\documentclass[twocolumn,aps,prc,superscriptaddress,showpacs,floatfix]{revtex4}

\usepackage{multirow}
\usepackage{epsfig}
\usepackage{amsmath}

\usepackage[normalem]{ulem}  
\usepackage[dvips]{color} 

\renewcommand\sout{\bgroup \color{red} \ULdepth=-.5ex \ULset}

\newcommand{\Ex}[2]{\ifmmode{#1\times10^{#2}}\else{$#1\times10^{#2}$}\fi}

\begin{document}

\title{Majorana meets Coxeter: \\
Non-Abelian Majorana Fermions and 
Non-Abelian Statistics}

\vspace{5mm}

\author{Shigehiro Yasui}
\affiliation{KEK Theory Center, Institute of Particle 
and Nuclear Studies,
High Energy Accelerator Research Organization (KEK),
1-1 Oho, Tsukuba, Ibaraki 305-0801, Japan}

\author{Kazunori Itakura} 
\affiliation{KEK Theory Center, Institute of Particle 
and Nuclear Studies,
High Energy Accelerator Research Organization (KEK),
1-1 Oho, Tsukuba, Ibaraki 305-0801, Japan}

\author{Muneto Nitta}
\affiliation{Department of Physics, and Research and Education Center 
for Natural Sciences, Keio University, 4-1-1 Hiyoshi, Yokohama, 
Kanagawa 223-8521, Japan}

\date{\today}

\begin{abstract}
We discuss statistics of vortices having zero-energy 
non-Abelian Majorana fermions inside them. Considering 
the system of multiple non-Abelian vortices, we derive 
a non-Abelian statistics that differs from 
the previously derived non-Abelian statistics.
The new non-Abelian statistics presented here is given by a tensor 
product of two different groups, namely the non-Abelian 
statistics obeyed by the Abelian Majorana fermions and 
the Coxeter group. The Coxeter group is a 
symmetric group related to the
symmetry of polytopes in a high-dimensional space.
As the simplest example, we consider the case in which 
a vortex contains three Majorana fermions that are 
mixed with each other under the $SO(3)$ transformations.
We concretely present the representation of the 
Coxeter group in our case and its geometrical 
expressions in the high-dimensional Hilbert 
space constructed from non-Abelian Majorana fermions.
\end{abstract}


\maketitle

\setcounter{page}{1}
\setcounter{footnote}{0}
\renewcommand{\thefootnote}{\arabic{footnote}}

\section{Introduction}

Majorana fermions proposed by Ettore Majorana in the 
20th century have a peculiar property in that an antiparticle 
is equivalent to a particle \cite{Majorana:1937vz}. 
His original conjecture that 
neutrinos might be such particles has been rejected; 
however, Majorana fermions are now attracting much attention in 
condensed matter physics \cite{Wilczek:2009}. 
It has been recently recognized that Majorana 
fermions with exact zero energy are trapped inside the 
core of half-quantized vortices in chiral $p$-wave 
superconductors or $p$-wave superfluids \cite{Volovik:1999} 
and that they 
appear on the edge of topological superconductors and insulators 
\cite{topological}. A system of multiple vortices provides 
an exchange statistics that differs from that of bosons, 
fermions and anyons. Such statistics, called ``non-Abelian 
statistics," is mathematically described by the braid 
group \cite{Read:1999fn,Ivanov:2001}. 
More precisely, one Dirac fermion is defined by a set of 
two Majorana fermions trapped in two different vortices, 
and the Hilbert space is constructed from these 
Dirac fermions \cite{Ivanov:2001}. The dimension $(2^{m})$ of 
the Hilbert space increases exponentially with an increase 
in the even number $(2m)$ of vortices. Consequently, non-Abelian 
statistics provides one possible candidate for quantum 
computation \cite{Nayak:2008zza}, and hence, $p$-wave 
superconductors or superfluids may 
be used as a device for 
quantum computers. This is why Majorana fermions and 
non-Abelian statistics have attracted so much attention 
in recent years \cite{Nayak:2008zza,Wilczek:2009}.
Recently, it has been proposed that even in three 
dimensions, non-Abelian statistics 
is realized 
by Majorana fermions trapped on monopole-like objects 
\cite{Teo:2009qv}. 

Though the exchange statistics 
discussed above is non-Abelian, 
Majorana fermions on vortices or other defects studied 
thus far are all ``Abelian" in the sense that only a 
single Majorana fermion is trapped in each defect. 
On the other hand, if multiple Majorana fermions are 
trapped inside a defect and can continuously mix with 
each other, one may be able to interpret them as a 
single ``non-Abelian Majorana fermion," namely 
a Majorana fermion having a non-Abelian 
internal symmetry. Then, these fermions must 
belong to representations of the underlying Lie group.

In this article, we show that non-Abelian Majorana 
fermions obey a non-Abelian exchange statistics 
that has not been identified before.
Since Majorana fermions are real ({\it i.e.} not complex) fields, 
their group representation must also be real. 
The simplest of such representation is the vector 
representation, triplet, of the $SO(3)$ Lie group. 
We explicitly construct the non-Abelian exchange statistics of 
non-Abelian Majorana fermions belonging to
the vector representation of $SO(3)$. 
We show not only that the Hilbert space for $2m$ vortices has 
a dimension $2^{3m}$ that is much larger than the $2^{m}$ 
of Abelian fermions but also that  
it contains a new component. 
In addition to the non-Abelian statistics already derived 
by Ivanov \cite{Ivanov:2001}, we find another structure, 
{\it i.e.} the Coxeter group \cite{Coxeter,book}: 
the entire non-Abelian statistics is a tensor product 
of these two. The Coxeter group is a symmetry group of 
higher-dimensional generalization of polytopes such as a
triangle or a tetrahedron, 
which was found by Harold Scott MacDonald (``Donald") Coxeter,
one of the great mathematicians of the 20th century.  
The large Hilbert space spanned by 
non-Abelian Majorana fermions 
contains high-dimensional internal spaces of various 
representations of $SO(3)$, not only singlets and triplets  
but also quintets and higher representations in general, 
where the Coxeter group acts on them to exchange 
multiple states in the same representations.

One question arising immediately may be whether 
there exist physical systems realizing such 
non-Abelian Majorana fermions in reality. 
The answer is yes, such systems probably exist in the 
universe, {\it i.e.}  in quark matter at extremely high density in
neutron stars or quark stars, 
which are expected to exhibit the so-called 
color superconductivity \cite{Alford:1997zt,Alford:2007xm}. 
These stars rotate rapidly, and consequently, 
stable vortices are created; these vortices are 
non-Abelian vortices with color magnetic fluxes 
confined inside them \cite{Balachandran:2005ev}.
We have shown in our previous paper \cite{Yasui:2010yw}
that non-Abelian Majorana fermions of an $SO(3)$ triplet 
 indeed exist in the core of a non-Abelian vortex.
The origin of this $SO(3)$ group is 
the color-flavor locked symmetry $SU(3)$ 
in the ground state of a color superconductor 
\cite{Alford:1997zt,Alford:2007xm},  
which is spontaneously broken down 
to its subgroup $SU(2) \times U(1)$ 
in the core of a vortex \cite{Nakano:2007dr}. 
Since there remains an unbroken symmetry 
$SU(2)\sim SO(3)$ inside the core, the zero modes trapped in it 
must belong to the representations of $SO(3)$. 
We have found triplet and singlet Majorana fermions 
\cite{Yasui:2010yw}; however, only the triplet 
is a new object to be considered in this article. 
Theoretically, fermionic modes in vortices 
are treated by the Bogoliubov-de Gennes equation, 
which is an equation for fermions coupled to the vortex 
profile. For example, the particle ($\varphi$) and hole 
($\eta$) components of the triplet zero modes 
(chirally right-handed)
are approximately given, in two-component Weyl spinor 
representation, as
\begin{eqnarray}
 \varphi(r,\theta) 
&=& {\rm e}^{-|\Delta_{\rm CFL}|r}
\left(
\begin{array}{c}
 J_{0}(\mu r) \\
 i J_{1}(\mu r)\, {\rm e}^{i\theta}
\end{array}
\right), \\ 
 \eta(r,\theta) &=& {\rm e}^{-|\Delta_{\rm CFL}|r}
\left(
\begin{array}{c}
 -J_{1}(\mu r)\, {\rm e}^{-i\theta} \\
 i J_{0}(\mu r) \\
\end{array}
\right), \label{tripletZMs}
\end{eqnarray}
where $\Delta_{\rm CFL}$ is the gap value of the 
bulk color superconductor (in the color-flavor locked phase), 
$\mu$ is the chemical potential, and $r,\theta$ are the polar coordinates 
perpendicular to the $z$ direction. The triplet zero modes $\psi^a\, 
(a=1,2,3)$ are then compactly expressed as 
$\psi^a \propto (\varphi, (-1)^{a+1} \eta)^{t}$, 
and they satisfy the Majorana condition $\psi^a=(\psi^a)^{C}$, 
where 
$(\psi^{a})^{C} \propto (\eta^{C},  (-1)^{a+1}\varphi^{C})^{t}$ 
with $C$ being the charge conjugation. 
It should be noted that 
these zero-mode solutions (non-Abelian Majorana fermions) are well localized 
around the center of 
the vortex and do not depend on the $z$ coordinate, indicating that 
they are essentially local objects on a two-dimensional plane.
Hence, the non-Abelian vortices appearing in high-density matter
provide an example of realization of the non-Abelian Majorana fermions.

However, it should be emphasized that 
our conclusion in this article does not rely on any specific 
model; all that we need is non-Abelian Majorana fermions of 
an $SO(3)$ triplet.  
Thus, our analysis raises a possibility to realize such 
a statistics in table top samples, for instance, in cold atomic 
gasses that can be well controlled through experiments.
It should also be noted that 
the $SO(3)$ group and its vector representation are chosen only
for illustration as the simplest example 
and that our method works for arbitrary Lie groups and 
arbitrary (real) representations, 
opening up a new possibility of 
Majorana fermions, non-Abelian statistics, 
and quantum computations.

This paper is organized as follows.
After recalling non-Abelian statistics 
of Abelian vortices in Sec.~\ref{sec2}, 
we construct the matrix representation of 
exchange operators of two and four 
non-Abelian vortices  
in Sec.~\ref{sec3}. 
We then identify the Coxter group 
from the decomposition of those matrices 
in Sec.~\ref{sec4}. 
Section \ref{sec5} is devoted to 
summary and discussion. 
In Appendix \ref{subsec1} we present explicit forms 
of the basis of the Hilbert spaces for 
four non-Abelian vortices.
In Appendix \ref{subsec2} we give a decomposition of 
exchange matrices at the operator level.

\section{Statistics on two-dimensional plane}\label{sec2}

\begin{figure}[tbp]
\begin{center}
\includegraphics[height=2.8cm,keepaspectratio,angle=0]{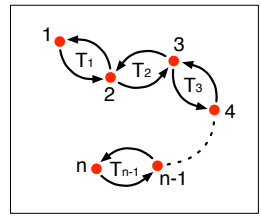}
\end{center}\vspace*{-5mm}
\caption{Schematic of $n$ particles on a two-dimensional 
plane and exchange of the $k$th and $(k+1)$th particles 
denoted by $T_{k}$ ($k=1$, $\cdots$, $n-1$).}
\label{fig:n_particles}
\end{figure}

\begin{figure}[tbp]
\begin{center}
\includegraphics[height=1.1in,keepaspectratio,angle=0]{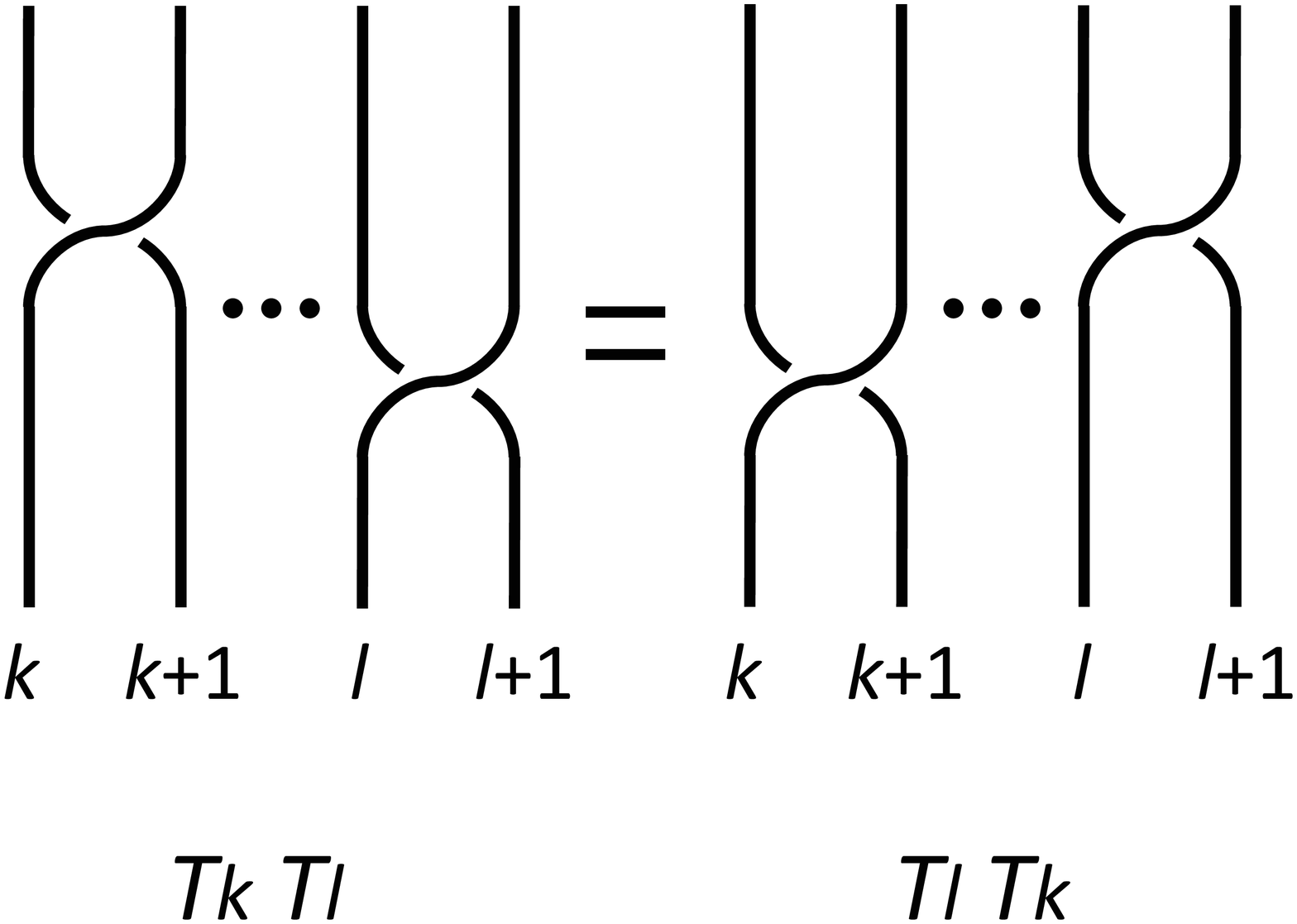}
\includegraphics[height=1.1in,keepaspectratio,angle=0]{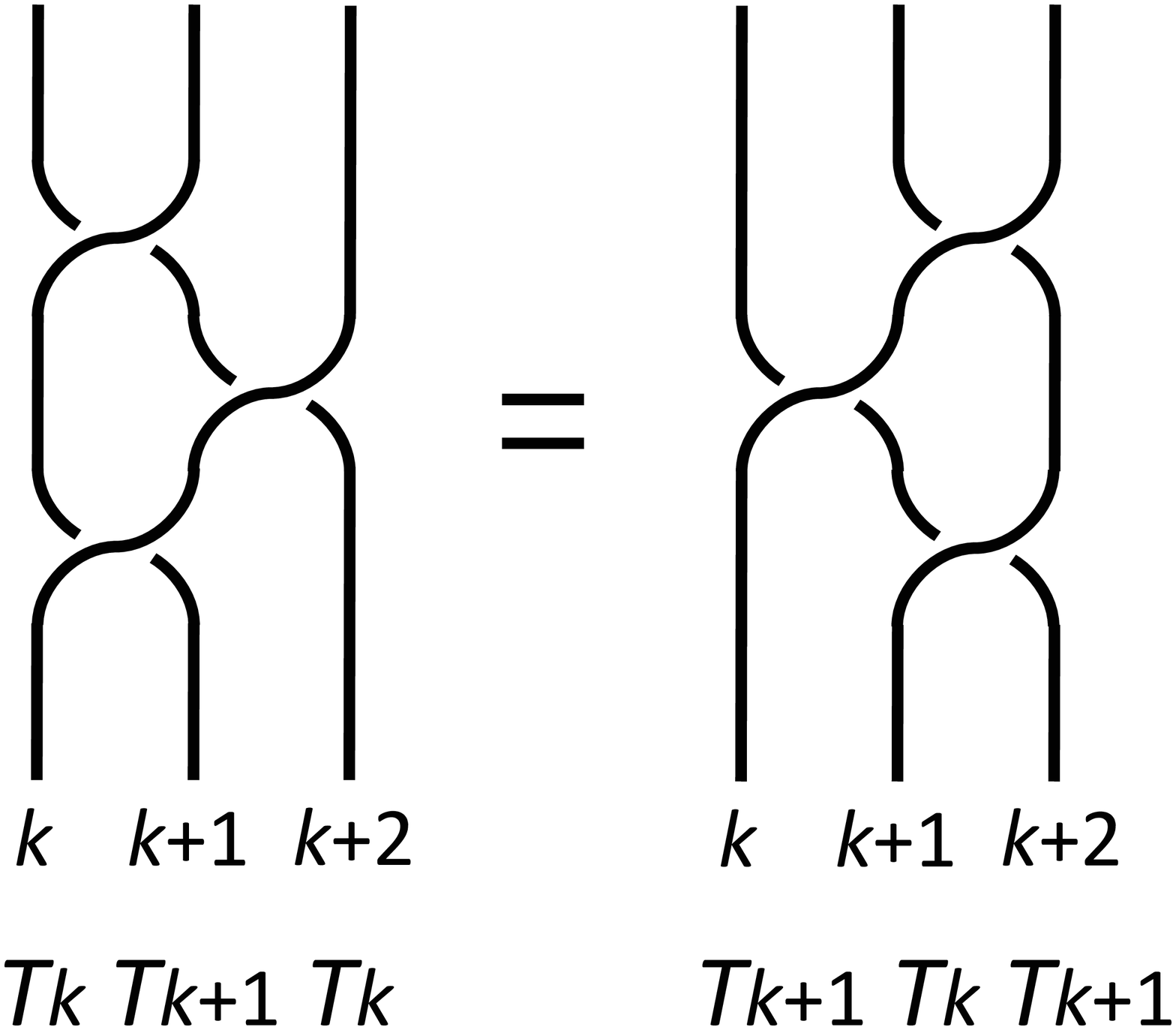}
\end{center}\vspace*{-5mm}
\caption{Schematic of the braid relations 
(i) $T_{k}T_{l}=T_{l}T_{k}$ with $|k-l|>1$ and 
(ii) $T_{k}T_{k+1}T_{k}=T_{k+1}T_{k}T_{k+1}$.}
\label{fig:TT}
\end{figure}

The exchange of particles on a two-dimensional plane is 
described by the braid group. 
Let us suppose $n$ particles (braids) and label them
as shown in Fig.~\ref{fig:n_particles}. 
The braid group is defined as a set of operations 
$T_{k}$ ($k=1$, $\cdots$, $n-1$) that involve the 
exchange of the positions of the neighboring $k$th and 
$(k+1)$th braids in a way such that
the $k$th braid always goes around the $(k+1)$th 
braid in an anti-clockwise direction.
The operations $T_k$ satisfy the following braid relations: 
(i) $T_k T_l = T_l T_k$ for $|k-l|>1$ and (ii) 
$T_k T_l T_k = T_l T_k T_l$ for $|k-l|=1$, which are schematically 
shown in Fig.~\ref{fig:TT}.
It should be noted that $T_k^{-1} \neq T_k$ 
because the operation is directed. 
This simple definition of the braid group allows for 
various representations with rich non-trivial structures, as we will
see below.

The representation of the braid group is expressed by a
linear group  $\{ \tau_k | k=1, \cdots, n-1\}$ 
acting on a vector space. Here, $\tau_k$'s also satisfy
the following braid relations: (i) $\tau_k \tau_l = \tau_l \tau_k$ for 
$|k-l|>1$ and (ii) $\tau_k \tau_l \tau_k = \tau_l \tau_k \tau_l$ 
for $|k-l|=1$. The representations contain 
information about the statistics on the exchange of particles.
For example, the one-dimensional representation allows for
the anyon statistics, which gives a wave function a complex factor 
under the exchange of particles; 
$\tau_1 = \tau _2 = \cdots = \tau_{n-1}={\rm e}^{i\theta}$ with 
$0\le \theta < 2\pi$ being a real number. 
Although the anyon statistics is a characteristic 
statistics in two dimensions, it is still Abelian.
More generally, the braid group allows for non-Abelian statistics 
in which neighboring $\tau_k$'s are noncommutative:
$[\tau_k, \tau_l] \neq 0$ for $|k-l|=1$.

\section{Non-Abelian statistics of multiple 
non-Abelian vortices }\label{sec3}

\begin{figure}[tbp]
\begin{center}
\includegraphics[height=2.4cm,keepaspectratio,angle=0]
{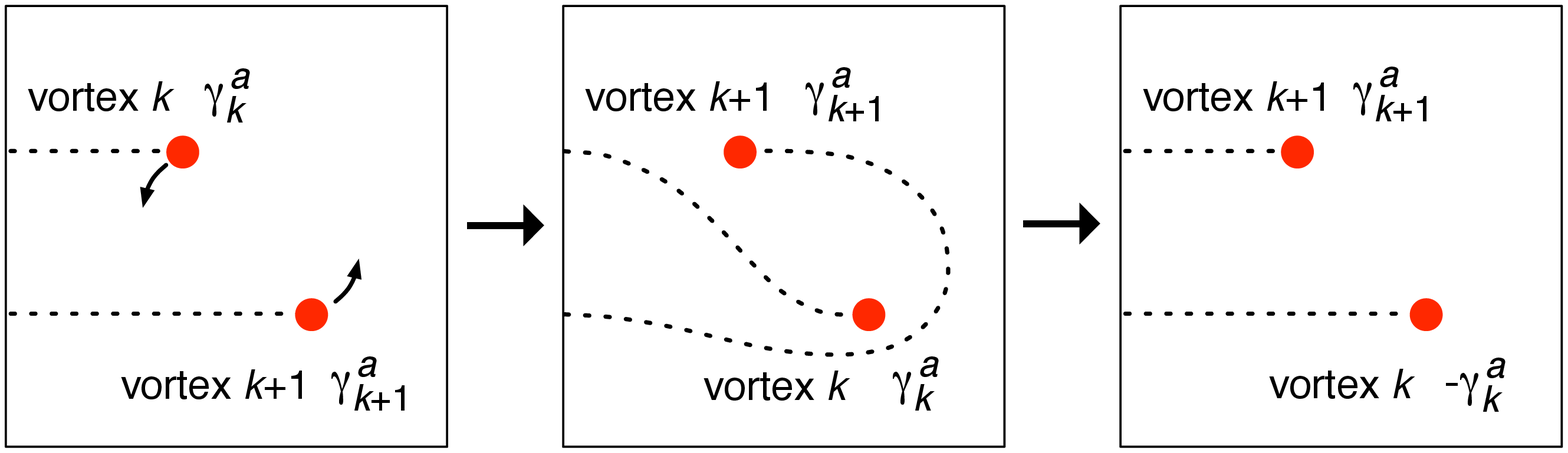}
\end{center}\vspace*{-5mm}
\caption{Exchange of two non-Abelian Majorana fermions 
$\gamma_{k}^{a}$ and $\gamma_{k+1}^{a}$ on the two-dimensional plane. 
The dotted lines denote the cuts attached to each non-Abelian Majorana 
fermion. From the middle figure to the right-most figure,
 it can be seen that 
the non-Abelian Majorana 
fermion $\gamma_k^a$ jumps the cut of the $(k+1)$th vortex 
to obtain an extra minus sign. Therefore, $T_{k}$ induces the 
transformation: $\gamma_{k}^{a} \rightarrow 
\gamma_{k+1}^{a}$ and $\gamma_{k+1}^{a} \rightarrow 
-\gamma_{k}^{a}$.}
\label{fig:exchange}
\end{figure}

Let us consider $n$ non-Abelian vortices with the $SO(3)$ symmetry.
There are correspondingly $n$ non-Abelian Majorana fermions 
belonging to the triplet of $SO(3)$ at each vortex site.
Let us introduce operators 
$\gamma_{k}^{a}$ ($a=1$, $2$, $3$) for creation operators of the 
triplet non-Abelian Majorana fermions in the $k$th non-Abelian 
vortex. These operators satisfy the anticommutation relation 
$\{\gamma_k^a , \gamma_l^b \} = 2 \delta_{kl} \delta^{ab}$ 
and the self-conjugation relation ${\gamma_{k}^{a}}^{\dag} = \gamma_{k}^{a}$. 
This formulation is an extension of the Abelian Majorana fermions 
discussed in Ref.~\cite{Ivanov:2001} to the non-Abelian case.
An exchange operation of neighboring $k$th and 
$(k+1)$th non-Abelian vortices, denoted by  
$T_k$ as shown in Fig.~\ref{fig:n_particles}, 
induces an exchange of the non-Abelian Majorana fermions. 
Because the Majorana fermion turning around a vortex 
changes the sign of the wave function owing to the odd winding 
number and the Majorana condition, a cut should be
introduced to keep the phase single valued \cite{Ivanov:2001}.
Consequently, the operation $T_k$ induces 
the following transformation (see Fig.~\ref{fig:exchange}):
\begin{eqnarray}
T_k : \left\{ 
\begin{array}{l}
 \gamma_{k}^{a}\quad \rightarrow\, \gamma_{k+1}^{a} \\
 \gamma_{k+1}^{a} \rightarrow -\gamma_{k}^{a}  
\end{array}
\right. , \quad {\rm for \ all}\ a
\label{eq:exchange_Majorana}
\end{eqnarray}
with the rest $\gamma_l^a$ ($l \neq k$, $k+1$) 
unchanged. It will turn out that the presence of 
this minus sign is crucial to induce the non-Abelian statistics.

Now, we discuss the representation $\tau_k$ for the exchange of 
non-Abelian Majorana fermions. 
First, it should be noted that transformation 
(\ref{eq:exchange_Majorana}) can be realized by $SO(3)$ invariant 
unitary operators $\hat \tau_k = \hat \tau_{k}^{1} \hat \tau_{k}^{2} 
\hat \tau_{k}^{3}$, where $\hat \tau_k^a$ is made of 
non-Abelian Majorana operators $\gamma_k^a$ and $\gamma_{k+1}^a$ 
as $\hat \tau_k^a = (1+\gamma_{k+1}^a \gamma_{k}^a)/\sqrt{2}$ 
(not summed over $a$). 
It is easily verified that $\hat \tau_k \gamma_l^a \hat \tau_k^{-1}$ 
indeed generates the above transformation.
Once the Hilbert space is defined, one obtains the 
representations of $\hat \tau_k$ as matrices. In order to 
define the bases of the Hilbert space, 
we first introduce an operator of non-Abelian Dirac fermions 
$\hat\Psi_k^a=(\gamma_{2k-1}^{a} +i \gamma_{2k}^{a})/2$ 
($k=1$, $\cdots$, $n/2$) using an even number $(n)$ of non-Abelian 
vortices. The non-Abelian Dirac fermions satisfy the 
anti-commutation relations 
$\{ \hat\Psi_k^a,  \hat \Psi_l^{b\dag} \}= \delta_{kl} \delta^{ab}$, 
$\{ \hat\Psi_k^a,  \hat \Psi_l^{b} \}=
 \{ \hat\Psi_k^{a\dag}, \hat \Psi_l^{b\dag} \}=0$,
and the operators $\hat\Psi_k^a$ and 
$\hat \Psi_l^{a\dag}(\neq \hat\Psi_k^a)$
correspond to annihilation and creation operators, respectively.
Then, we can construct the Hilbert space by acting successively 
creation operators $\hat\Psi_{k}^{a\dag}$'s 
on the ``vacuum state" $|0\rangle$ defined by 
$\hat\Psi_k^a |0\rangle = 0$.
In what follows, we concretely show representations $\tau_k$ in 
two cases with two ($n=2$) and four ($n=4$) non-Abelian vortices.

First, we discuss the case with two non-Abelian 
vortices ($n=2$), where we can define only one operation $T_1$
and only one non-Abelian Dirac fermion $\hat\Psi_1^a$. 
The Hilbert space is spanned by the following four basis states:
singlet-even $| {\bf 1}_{0} \rangle 
            = | 0 \rangle$ (vacuum), 
singlet-odd  $| {\bf 1}_{3} \rangle = 
           \frac{1}{3!}\epsilon^{abc}
           \hat \Psi_1^{a\dag} \hat \Psi_1^{b\dag}\hat \Psi_1^{c\dag} 
           | 0 \rangle$  
          (filled by three Dirac fermions), 
triplet-even $| {\bf 3}_{2} \rangle =
           \frac{1}{2!}\epsilon^{abc}
           \hat \Psi_1^{b\dag}\hat \Psi_1^{c\dag} 
           | 0 \rangle$ 
          (occupied by two Dirac fermions), 
and triplet-odd $| {\bf 3}_{1} \rangle =
           \hat \Psi_1^{a\dag} | 0 \rangle$
          (occupied by one Dirac fermion), 
where the subscript denotes the number of non-Abelian Dirac fermions.
With these bases 
\{$| {\bf 1}_{0} \rangle$, $| {\bf 1}_{3} \rangle$, 
$| {\bf 3}_{2} \rangle$, $| {\bf 3}_{1} \rangle$\}, we 
find the representation $\tau_1$ to be an 
$8 \times 8$ matrix; 
$\tau_{1} = \mbox{diag} ({\rm e}^{-i3\pi/4},\, {\rm e}^{i3\pi/4},\, 
{\rm e}^{i\pi/4} I_{3\times3},\, {\rm e}^{-i\pi/4} I_{3\times3})$.
The $3 \times 3$ unit matrix $I_{3\times 3}$ means the three 
components in each triplet state. 
The obtained matrix $\tau_1$ is diagonal and, thus, 
the system of two non-Abelian Majorana fermions follows 
the (anyonlike) Abelian statistics.

Second, we discuss the representation $\tau_{k}$ ($k=1$, $2$, 3) 
for four non-Abelian vortices ($n=4$).
We can construct the Hilbert space in a manner similar to 
that described above, and 
this is expressed by singlet (${\bf 1}$), 
triplet (${\bf 3}$), and quintet (${\bf 5}$) states, which are
further specified by even (${\cal E}$) and odd (${\cal O}$) 
numbers of Dirac fermions 
(see Appendix~\ref{subsec1}). Then,  
we obtain the representations $\tau_{k}$ ($k=1$, $2$, $3$) as 
$64 \times 64$ matrices; 
$\tau_{k} = \mbox{diag} 
(\tau_{k}^{{\bf 1}, {\cal E}},\, 
 \tau_{k}^{{\bf 1}, {\cal O}},\, 
 \tau_{k}^{{\bf 3}, {\cal E}} I_{3\times3},\, 
 \tau_{k}^{{\bf 3}, {\cal O}} I_{3\times3},\, 
 \tau_{k}^{{\bf 5}, {\cal E}} I_{5\times5},\, 
 \tau_{k}^{{\bf 5}, {\cal O}} I_{5\times5})$.
The submatrices $\tau_{k}^{{\cal M}, {\cal P}}$ 
(${\cal M}={\bf 1}$, ${\bf 3}$, ${\bf 5}$, 
 ${\cal P}={\cal E}$, ${\cal O}$)
are found to have the following structure:
\begin{eqnarray}
\tau_{k}^{{\cal M}, {\cal P}} = 
\sigma_{k}^{{\cal M}} \otimes h_{k}^{{\cal P}}\, .
\label{eq:sigma_h}
\end{eqnarray}
Here, the first matrices $\sigma_{k}^{{\cal M}}$ have 
different expressions depending on the multiplets: 
\begin{eqnarray}
\sigma_{1}^{{\bf 1}} =
\left(
\begin{array}{cc}
 -1 & 0 \\
 0 & 1
\end{array}
\right),\quad 
\sigma_{2}^{{\bf 1}} = \frac{1}{2}
\left(
\begin{array}{cc}
 1 & \sqrt{3} \\
 \sqrt{3} & -1
\end{array}
\right),\quad 
\sigma_{3}^{{\bf 1}} = \sigma_{1}^{{\bf 1}}
 ,
\label{eq:sigma1}
\end{eqnarray}
for the singlet states, 
\begin{eqnarray}
&&\sigma_{1}^{{\bf 3}} =
\left(
\begin{array}{ccc}
 -1 & 0 & 0 \\
 0 & 1 & 0 \\
 0 & 0 & 1
\end{array}
\right)\! , \ 
\sigma_{2}^{{\bf 3}} = \frac{1}{2}\!
\left(
\begin{array}{ccc}
 1 & \sqrt{2} & 1 \\
 \sqrt{2} & 0 & -\sqrt{2} \\
 1 & -\sqrt{2} & 1
\end{array}
\right)\! ,\quad \nonumber \\
&&\sigma_{3}^{{\bf 3}} =
\left(
\begin{array}{ccc}
 1 & 0 & 0 \\
 0 & 1 & 0 \\
 0 & 0 & -1
\end{array}
\right)\! ,
\label{eq:sigma3}
\end{eqnarray}
for the triplet states, and 
\begin{eqnarray}
\sigma_{1}^{{\bf 5}} = \sigma_{2}^{{\bf 5}} = \sigma_{3}^{{\bf 5}} = 1,
\label{eq:sigma5}
\end{eqnarray}
for the quintet states. On the other hand, 
the second matrices $h_{k}^{{\cal P}}$ are common for the multiplets: 
\begin{eqnarray}
&&\hspace*{-4mm} h_{1}^{{\cal E}}\! =\! h_{1}^{{\cal O}}\! =\!
\left(
\begin{array}{cc}
 {\rm e}^{i\frac{\pi}{4}} & 0 \\
 0 & {\rm e}^{-i\frac{\pi}{4}}
\end{array}
\right)\! ,\ 
h_{2}^{{\cal E}}\! =\! h_{2}^{{\cal O}}\! =\! \frac{1}{\sqrt{2}}\!
\left(
\begin{array}{cc}
 1 & -1 \\
 1 & 1
\end{array}
\right)\! ,\quad \nonumber \\
&&\hspace*{-4mm} h_{3}^{{\cal E}}\! =\! h_{3}^{{\cal O}\dag} 
\!=\! h_{1}^{{\cal E}} 
\label{eq:h}
\end{eqnarray}
for even and odd numbers of Dirac fermions. 
These are our main results.
We have obtained matrix representations for the exchange 
statistics in the system of four non-Abelian vortices. 
It should be noted that both the matrices $\sigma_2^{\cal M}\, 
({\cal M}={\bf 1,3})$ and $h^{\cal P}_2$ in 
the submatrices $\tau_2^{{\cal M},{\cal P}}$, and consequently
the matrix $\tau_2$, are nondiagonal. 
Therefore, {\it the system of four non-Abelian Majorana 
fermions follows the non-Abelian statistics.} It should be 
emphasized that 
the non-Abelian matrices we have derived are essentially new 
and are 
considered as generalizations of the corresponding matrices 
obtained by Ivanov.
Indeed, while the matrices $h_{k}^{{\cal P}}$ (common for the 
multiplets) are the same as those that Ivanov obtained for 
``Abelian" vortices (thus we call them the Ivanov matrices),
the matrices $\sigma_{k}^{{\cal M}}$ are 
new matrices that appear only in ``non-Abelian" vortices.
Hence, the representation matrices we have found are the 
tensor products of the
new matrices $\sigma_{k}^{{\cal M}}$ and Ivanov matrices 
$h_{k}^{{\cal P}}$.

\section{Coxeter group }\label{sec4}

Unexpectedly, the new matrices $\sigma_{k}^{{\cal M}}$ 
for the four non-Abelian vortices 
are identified with the elements in 
the Coxeter group, which is related to the 
operations of reflections shown below. 
The Coxeter group $S$ is defined as a group with distinct
generators $s_i \in S$ ($i=1,2,3,\cdots$)
satisfying the following two conditions: 
(a) $s_i^{2}=1$ and (b) $(s_i\,s_j)^{m_{i,j}}=1$ 
with a positive integer $m_{i,j} \ge 2$ for $i \neq j$.
It should be noted that condition (a) gives $m_{i,i}=1$.
Elements $m_{i,j}$ can be summarized as 
the Coxeter matrix $({\sf M})_{ij}=m_{i,j}$.
It is easy to check that the matrices 
$\sigma_{k}^{{\cal M}}$ indeed satisfy conditions (a) and (b),
and hence $\sigma_{k}^{{\cal M}}$ can be regarded as generators of the Coxeter group.
Therefore, 
we obtain the Coxeter matrix for the triplet [Eq.~(\ref{eq:sigma3})] 
as
\begin{eqnarray}
{\sf M}_3 = \left(
\begin{array}{ccc}
 1 & 3 & 2 \\
 3 & 1 & 3 \\
 2 & 3 & 1
\end{array}
\right).
\label{eq:Coxeter3}
\end{eqnarray}
The Coxeter matrices for the other representations are given by submatrices in ${\sf M}_3$.
The Coxeter matrix for the singlet is the top-left (bottom-right) $2 \times 2$ submatrix 
$\left(\begin{array}{cc}
      1 &  3 \\
      3 & 1
\end{array}\right)$
in ${\sf M}_3$, because there are only two 
distinct operators $\sigma_{1}^{{\bf 1}}$ and $\sigma_{2}^{{\bf 1}}$ 
($\sigma_{2}^{{\bf 1}}$ and $\sigma_{3}^{{\bf 1}}$) in 
Eq.~(\ref{eq:sigma1}). 
The Coxeter matrix for the quintet is the top-left (middle or 
bottom-right) $1 \times 1$ submatrix $(1)$ in ${\sf M}_3$, 
because there is only one distinct operator $\sigma_{1}^{{\bf 5}}$ 
($\sigma_{2}^{{\bf 5}}$ or $\sigma_{3}^{{\bf 5}}$) 
in Eq.~(\ref{eq:sigma5}).

\begin{figure}[tbp]
\begin{center}
\includegraphics[height=4.5cm,keepaspectratio,angle=0]{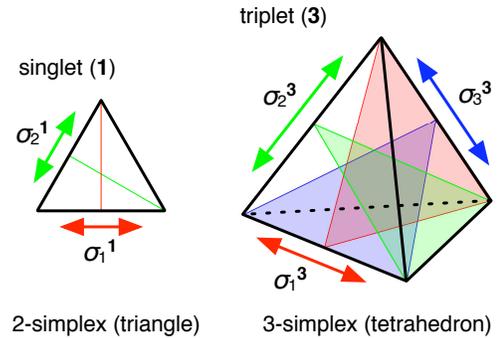}
\end{center}\vspace*{-3mm}
\caption{2-simplex (triangle) for the singlet and 3-simplex (tetrahedron) for the triplet induced from the reflections $\sigma_{k}^{{\cal M}}$ ($k=1$ and $2$ for ${\cal M}={\bf 1}$, and $k=1$, $2$, and $3$ for ${\cal M}={\bf 3}$). These simplexes are invariant under reflections by $\sigma_{k}^{{\cal M}}$.}
\label{fig:triangle_tetrahedron}
\end{figure}

The Coxeter group is 
closely related to the geometry in a high-dimensional space,
because conditions (a) and (b) can be 
interpreted as geometrical operations in the Hilbert space: 
Condition (a) corresponds to a reflection and 
condition (b) corresponds to 
a rotation by an angle 
$2\pi/m_{i,j}$. 
Therefore, the Coxeter group leads to the existence 
of several polytopes that are invariant under reflections 
and rotations,
such as a 2-simplex (triangle) under the reflections $\sigma_{1}^{\bf 1}$ 
and $\sigma_{2}^{\bf 1}$ for the singlet and a 3-simplex (tetrahedron) 
under the reflections $\sigma_{1}^{\bf 3}$, $\sigma_{2}^{\bf 3}$, and 
$\sigma_{3}^{\bf 3}$ for the triplet, as shown in 
Fig.~\ref{fig:triangle_tetrahedron}. 
The fact that the Coxeter group appears in the 
exchange statistics of the system of the non-Abelian vortices
provides a new insight that the non-Abelian statistics of 
non-Abelian Majorana fermions can be intuitively understood 
with the help of the geometry of polytopes. 

We may extend our discussion to the system of any even number 
$(n=2m)$ of non-Abelian vortices. 
Because an interchange of vortices does not mix the $i$th and 
$j$th vortices for $|i-j|>1$, the essential structure is already 
determined from the analysis for four vortices.
Therefore, we find $m_{i,j}=3$ when $|i-j|=1$ and 
$m_{i,j}=2$ when $|i-j| > 1$ for the 
representation matrices $\sigma_{i}$ ($i=1, \cdots, 2m-1$), 
which are obtained from the decomposition of $\tau_{k}$.
Hence, the Coxeter matrix will be 
given by the $(2m-1)\times(2m-1)$ matrix as follows: 
\begin{eqnarray}
{\sf M}_{2m-1} = \left(
\begin{array}{ccccc}
 1 & 3 & 2 & 2 & \cdots\\
 3 & 1 & 3 & 2 & \\
 2 & 3 & 1 & 3 & \\
 2 & 2 & 3 & 1 &  \\
 \vdots & & & & \ddots 
\end{array}
\right).
\label{eq:Coxeter}
\end{eqnarray}
In other words,
a product $\sigma_{i}\sigma_{j}$ of reflections $\sigma_{i}$ and 
$\sigma_{j}$ makes a rotation by an angle $2\pi/3$ for $|i-j|=1$ 
and $2\pi/2=\pi$ for $|i-j| > 1$.
Therefore, the system of an even number $(2m)$ of non-Abelian vortices 
leads to the existence of a $(2m-1)$-simplex 
as the highest-dimensional object
 (see also Appendix~\ref{subsec2}).
The Coxeter group summarized by the matrix (\ref{eq:Coxeter}) 
is called $A_{2m-1}$, 
which is in fact 
isomorphic to the permutation group of $2m$ elements.

\section{Suammary and Discussion }\label{sec5}
In conclusion, we have explicitly constructed the 
non-Abelian statistics, described by the Coxeter 
group, that appears in a system of non-Abelian vortices 
having zero-energy non-Abelian Majorana fermions. 
Although color superconductors are only one example realizing 
such non-Abelian vortices thus far, our result is robust: 
Any system admitting vortices with three zero-energy 
Majorana fermions will have this non-Abelian statistics.
Cold atomic gasses such as Bose-Fermi mixture may 
offer such examples. We hope that not only the large 
dimension of the Hilbert space but also the Coxeter 
group should play important roles in quantum computations.

It is known that the Coxeter group is classified into 
several types \cite{book}. 
As denoted above, the Coxeter group summarized by the 
matrix (\ref{eq:Coxeter}) is called $A_{2m-1}$. 
Then, it would be interesting to ask whether the system of 
non-Abelian Majorana fermions can lead to other types of 
Coxeter groups, such as 
$B_{l}$ ($l\ge2$), $D_{l}$ ($l\ge4$), $E_{l}$ ($l=6$, $7$, $8$), 
$F_{4}$, $G_{2}$, $H_{l}$ ($l=3$, $4$), and $I_{2}(l)$ ($l=5$, $l\ge 7$), 
or which new symmetry of the non-Abelian Majorana fermions 
would be connected to them.
These questions would open an extensive view about the  
relationship between the non-Abelian Majorana fermions and 
the Coxeter groups.

\section*{Acknowledgments}
This work is supported in part by a Grant-in-Aid for Scientific Research on 
Priority Areas ``Elucidation of New Hadrons with a Variety of Flavors 
(E01: 21105006)" (S. Y.) and by Grant-in-Aid for Scientific Research No. 20740141 (M. N.) from 
the ministry of Education, Culture, Sports, Science and Technology of Japan.

\begin{appendix}

\section{Bases of the Hilbert space for the system of 
four non-Abelian vortices}\label{subsec1}

We explicitly  present  the bases of the Hilbert space 
of four non-Abelian vortices.
There are four bases,
\begin{eqnarray}
&&\hspace{-1em} |{\bf 1}_{00}\rangle = |0\rangle, \\
&&\hspace{-1em} |{\bf 1}_{33}\rangle 
   = i \frac{1}{3!} \epsilon^{abc} \frac{1}{3!} \epsilon^{def} 
     \hat{\Psi}_{1}^{a\dag} 
     \hat{\Psi}_{1}^{b\dag} 
     \hat{\Psi}_{1}^{c\dag} 
     \hat{\Psi}_{2}^{d\dag} 
     \hat{\Psi}_{2}^{e\dag} 
     \hat{\Psi}_{2}^{f\dag}  |0\rangle, \\
&&\hspace{-1em} |{\bf 1}_{11}\rangle 
   = i\frac{1}{\sqrt{3}} 
     \hat{\Psi}_{1}^{a\dag} 
     \hat{\Psi}_{2}^{a\dag} |0\rangle, \\
&&\hspace{-1em} |{\bf 1}_ {22}\rangle 
   = \frac{1}{\sqrt{3}} \frac{1}{2!} 
     \epsilon^{abc} \frac{1}{2!} \epsilon^{ade} 
     \hat{\Psi}_{1}^{b\dag} 
     \hat{\Psi}_{1}^{c\dag} 
     \hat{\Psi}_{2}^{d\dag} 
     \hat{\Psi}_{2}^{e\dag} |0\rangle,  
\end{eqnarray}
for the singlet-even (${\bf 1}$, $\cal{E}$) states and
\begin{eqnarray}
&&\hspace{-1em} |{\bf 1}_ {03}\rangle 
  = \frac{1}{3!} \epsilon^{abc} 
    \hat{\Psi}_{2}^{a\dag} 
    \hat{\Psi}_{2}^{b\dag} 
    \hat{\Psi}_{2}^{c\dag} |0\rangle, \\
&&\hspace{-1em} |{\bf 1}_{30}\rangle 
  = -i \frac{1}{3!} \epsilon^{abc} 
    \hat{\Psi}_{1}^{a\dag} 
    \hat{\Psi}_{1}^{b\dag} 
    \hat{\Psi}_{1}^{c\dag} |0\rangle, \\
&&\hspace{-1em} |{\bf 1}_{21}\rangle 
  = -\frac{1}{\sqrt{3}} \frac{1}{2!} \epsilon^{abc} 
    \hat{\Psi}_{1}^{a\dag} 
    \hat{\Psi}_{1}^{b\dag} 
    \hat{\Psi}_{2}^{c\dag} |0\rangle, \\
&&\hspace{-1em} |{\bf 1}_{12}\rangle 
  = i \frac{1}{\sqrt{3}} \frac{1}{2!} \epsilon^{abc} 
    \hat{\Psi}_{1}^{a\dag} 
    \hat{\Psi}_{2}^{b\dag} 
    \hat{\Psi}_{2}^{c\dag} |0\rangle, 
\end{eqnarray}
for the singlet-odd (${\bf 1}$, $\cal{O}$) states.
There are six bases,
\begin{eqnarray}
&&\hspace{-1.5em} |{\bf 3}_{02}\rangle 
  = \frac{1}{2!} \epsilon^{abc} 
    \hat{\Psi}_{2}^{b\dag} 
    \hat{\Psi}_{2}^{c\dag} |0\rangle, \\
&&\hspace{-1.5em} |{\bf 3}_ {31}\rangle 
  = -i \frac{1}{3!} \epsilon^{bcd} 
    \hat{\Psi}_{1}^{b\dag} 
    \hat{\Psi}_{1}^{c\dag} 
    \hat{\Psi}_{1}^{d\dag} 
    \hat{\Psi}_{2}^{a\dag} |0\rangle, \\
&&\hspace{-1.5em} |{\bf 3}_ {22}\rangle 
  = \frac{1}{\sqrt{2}} \epsilon^{abc} \frac{1}{2!} \epsilon^{bde} 
    \frac{1}{2!} \epsilon^{cfg} 
    \hat{\Psi}_{1}^{d\dag} 
    \hat{\Psi}_{1}^{e\dag}  
    \hat{\Psi}_{2}^{f\dag} 
    \hat{\Psi}_{2}^{g\dag} |0\rangle, \\ 
&&\hspace{-1.5em} |{\bf 3}_{11}\rangle 
  = i \frac{1}{\sqrt{2}} \epsilon^{abc} 
    \hat{\Psi}_{1}^{b\dag} 
    \hat{\Psi}_{2}^{c\dag}  |0\rangle, \\
&&\hspace{-1.5em} |{\bf 3}_{20}\rangle 
  = -\frac{1}{2!} \epsilon^{abc} 
    \hat{\Psi}_{1}^{b\dag} 
    \hat{\Psi}_{1}^{c\dag} |0\rangle, \\
&&\hspace{-1.5em} |{\bf 3}_ {13}\rangle 
  = i \frac{1}{3!} \epsilon^{bcd} 
    \hat{\Psi}_{1}^{a\dag} 
    \hat{\Psi}_{2}^{b\dag} 
    \hat{\Psi}_{2}^{c\dag} 
    \hat{\Psi}_{2}^{d\dag} |0\rangle, 
\end{eqnarray}
for the triplet-even (${\bf 3}$, $\cal{E}$) states and
\begin{eqnarray}
&&\hspace{-1em} |{\bf 3}_{01}\rangle 
  = \hat{\Psi}_{2}^{a\dag} |0\rangle, \\
&&\hspace{-1em} |{\bf 3}_ {32}\rangle 
  = i \frac{1}{3!} \epsilon^{bcd} 
    \hat{\Psi}_{1}^{b\dag} 
    \hat{\Psi}_{1}^{c\dag} 
    \hat{\Psi}_{1}^{d\dag} 
    \frac{1}{2!} \epsilon^{aef} 
    \hat{\Psi}_{2}^{e\dag} 
    \hat{\Psi}_{2}^{f\dag} |0\rangle, \\
&&\hspace{-1em} |{\bf 3}_ {21}\rangle 
  = \frac{1}{\sqrt{2}} \epsilon^{abc} \frac{1}{2!} \epsilon^{bde} 
    \hat{\Psi}_{1}^{d\dag} 
    \hat{\Psi}_{1}^{e\dag}  
    \hat{\Psi}_{2}^{c\dag} |0\rangle, \\
&&\hspace{-1em} |{\bf 3}_{12}\rangle 
  = -i \frac{1}{\sqrt{3}} \epsilon^{abc} \frac{1}{2!} \epsilon^{cde} 
    \hat{\Psi}_{1}^{b\dag} 
    \hat{\Psi}_{2}^{d\dag} 
    \hat{\Psi}_{2}^{e\dag} |0\rangle, \\
&&\hspace{-1em} |{\bf 3}_{23}\rangle 
  = \frac{1}{2!} \epsilon^{abc} 
    \hat{\Psi}_{1}^{b\dag} 
    \hat{\Psi}_{1}^{c\dag} 
    \frac{1}{3!} \epsilon^{def} 
    \hat{\Psi}_{2}^{d\dag} 
    \hat{\Psi}_{2}^{e\dag} 
    \hat{\Psi}_{2}^{f\dag} |0\rangle, \\
&&\hspace{-1em} |{\bf 3}_{10}\rangle 
  = i \hat{\Psi}_{1}^{a \dag} |0\rangle, 
\end{eqnarray}
for the triplet-odd (${\bf 3}$, $\cal{O}$) states.
There are two bases,
\begin{eqnarray}
&&\hspace{-1em} |{\bf 5}_{22}\rangle
  = i {\cal N} 
    \left\{\frac{1}{2} \left( \frac{1}{2!} \epsilon^{acd} 
                               \hat{\Psi}_{1}^{c\dag} 
                               \hat{\Psi}_{1}^{d\dag} 
                              \frac{1}{2!} \epsilon^{bef} 
                               \hat{\Psi}_{2}^{e\dag} 
                               \hat{\Psi}_{2}^{f\dag}      \right. \right. \nonumber \\
&&\hspace{7em}  + \left. \frac{1}{2!} \epsilon^{bcd} 
                               \hat{\Psi}_{1}^{c\dag} 
                               \hat{\Psi}_{1}^{d\dag} 
                              \frac{1}{2!} \epsilon^{aef} 
                               \hat{\Psi}_{2}^{e\dag} 
                               \hat{\Psi}_{2}^{f\dag} 
                      \right)
     \nonumber \\
&&\hspace{4em} 
     \left. - \frac{\delta^{ab}}{3} \frac{1}{2!} \epsilon^{cde} 
               \hat{\Psi}_{1}^{d\dag} 
               \hat{\Psi}_{1}^{e\dag} 
              \frac{1}{2!} \epsilon^{cfg} 
               \hat{\Psi}_{2}^{f\dag} 
               \hat{\Psi}_{2}^{g\dag} 
     \right\} |0\rangle, \\
&&\hspace{-1em} |{\bf 5}_{11}\rangle 
  = -{\cal N} 
     \left\{ \frac{1}{2} \left( \hat{\Psi}_{1}^{a\dag} 
                                \hat{\Psi}_{2}^{b\dag}    
                              + \hat{\Psi}_{1}^{b\dag} 
                                \hat{\Psi}_{2}^{a\dag} 
                        \right)
 - \frac{\delta^{ab}}{3} \hat{\Psi}_{1}^{c\dag} 
                                  \hat{\Psi}_{2}^{c\dag} 
     \right\} |0\rangle, \nonumber \\
\end{eqnarray}
for the quintet-even (${\bf 5}$, $\cal{E}$) states and
\begin{eqnarray}
&&\hspace{-1em} |{\bf 5}_{21}\rangle 
  = -i {\cal N} 
    \left\{ \frac{1}{2} \left( \frac{1}{2!} \epsilon^{acd} 
                                 \hat{\Psi}_{1}^{c\dag} 
                                 \hat{\Psi}_{1}^{d\dag} 
                                 \hat{\Psi}_{2}^{b\dag} 
+ \frac{1}{2!} \epsilon^{bcd} 
                                 \hat{\Psi}_{1}^{c\dag} 
                                 \hat{\Psi}_{1}^{d\dag} 
                                 \hat{\Psi}_{2}^{a\dag} 
                       \right) \right. \nonumber \\
&&\hspace{5em} \left. - \frac{\delta^{ab}}{3} \frac{1}{2!} \epsilon^{cde} 
               \hat{\Psi}_{1}^{c\dag} 
               \hat{\Psi}_{1}^{d\dag} 
               \hat{\Psi}_{2}^{e\dag} 
     \right\} |0\rangle, \\
&&\hspace{-1em} |{\bf 5}_{12}\rangle 
  = -{\cal N} 
    \left\{ \frac{1}{2} \left(   \hat{\Psi}_{1}^{a\dag} 
                                \frac{1}{2!}\epsilon^{bcd} 
                                 \hat{\Psi}_{2}^{c\dag} 
                                 \hat{\Psi}_{2}^{d\dag}
+ \hat{\Psi}_{1}^{b\dag} 
                                \frac{1}{2!} \epsilon^{acd} 
                                 \hat{\Psi}_{2}^{c\dag} 
                                 \hat{\Psi}_{2}^{d\dag} 
                       \right) \right. \nonumber \\
&&\hspace{5em} \left.  - \frac{\delta^{ab}}{3} \frac{1}{2!} \epsilon^{cde}
                    \hat{\Psi}_{1}^{c\dag} 
                    \hat{\Psi}_{2}^{d\dag} 
                    \hat{\Psi}_{2}^{e\dag}
    \right\} |0\rangle, 
\end{eqnarray}
for the quintet-odd (${\bf 5}$, $\cal{O}$) states with 
${\cal N}=\sqrt{3/2}$ for $a=b$ and ${\cal N}=\sqrt{2}$ for $a\neq b$.
With these bases we obtain the matrices $\tau_{k}$ ($k=1$, $2$, $3$) 
as presented in Eqs. (\ref{eq:sigma_h})--(\ref{eq:h}).

\section{Coxeter matrix for arbitrary number of non-Abelian vortices}\label{subsec2}

In the text, we show the tensor-product structure of the matrices 
$\tau_{k} = \sigma_{k}^{{\cal M}} \otimes h_{k}^{{\cal P}}$ 
(${\cal M}={\bf 1}$, ${\bf 3}$, ${\bf 5}$, and 
${\cal P}={\cal E}$, ${\cal O}$) obtained in the Hilbert space 
with bases presented in Appendix \ref{subsec1}. 
In fact, such a product structure holds even at the operator level.
The operator 
$\hat \tau_k = \hat \tau_{k}^{1} \hat \tau_{k}^{2} \hat \tau_{k}^{3}$ 
with $\hat \tau_k^a = (1+\gamma_{k+1}^a \gamma_{k}^a)/\sqrt{2}$ 
is expressed as a product of two $SO(3)$ invariant unitary 
operators, {\it i.e.} 
\begin{eqnarray}
\hat{\tau}_{k} = \hat{\sigma}_{k} \hat{h}_{k}\, ,
\end{eqnarray}
where
\begin{eqnarray}
&&\hat{\sigma}_{k} 
  = \frac{1}{2} 
    \Big( 1 - \gamma_{k+1}^{1}\gamma_{k+1}^{2}\gamma_{k}^{1}\gamma_{k}^{2} 
            - \gamma_{k+1}^{2}\gamma_{k+1}^{3}\gamma_{k}^{2}\gamma_{k}^{3} 
\nonumber \\
&&\hspace{4.5em} 
- \gamma_{k+1}^{3}\gamma_{k+1}^{1}\gamma_{k}^{3}\gamma_{k}^{1}
     \Big)
\end{eqnarray}
and
\begin{eqnarray}
\hat{h}_{k} 
  = \frac{1}{\sqrt{2}} 
    \Big( 1- \gamma_{k+1}^{1}\gamma_{k+1}^{2}\gamma_{k+1}^{3}
             \gamma_{k}^{1} \gamma_{k}^{2} \gamma_{k}^{3}
    \Big).
\end{eqnarray}
It is easily verified that operators $\hat{\sigma}_{k}$ 
satisfy relations (a) and (b) of the Coxeter group, such that
\begin{eqnarray}
\hat{\sigma}_{k}^{2} &=& 1, \nonumber \\
(\hat{\sigma}_{k} \hat{\sigma}_{l})^{3} &=& 1 \quad {\rm for}\quad |k-l|=1,
  \nonumber \\
(\hat{\sigma}_{k} \hat{\sigma}_{l})^{2} &=& 1 \quad {\rm for}\quad |k-l|>1.
\end{eqnarray}
Therefore, we confirm the Coxeter matrix (\ref{eq:Coxeter}) 
for an arbitrary number of non-Abelian vortices.

\end{appendix}


\end{document}